\documentclass[12pt]{iopart}

\usepackage{iopams}  
\usepackage{graphicx}
\usepackage{xcolor}
\expandafter\let\csname equation*\endcsname\relax
\expandafter\let\csname endequation*\endcsname\relax
 \usepackage{amsmath}
\usepackage{hyperref}
\usepackage{siunitx}
\usepackage{comment}
\begin{document}
\title{A New Design for a Traveling-Wave Zeeman Decelerator: I. Theory}

\author{Tomislav Damjanovi\'c$^{1}$, Stefan Willitsch$^{1}$, Nicolas Vanhaecke$^{2,3,\ddagger}$, Henrik Haak$^{2}$, Gerard Meijer$^{2}$, JeanPaul Cromi\'eres$^{3}$, Dongdong Zhang$^{*,4,1,2}$}
 \address{$^{1}$ 
 Department of chemistry, University of Basel, Klingelbergstrasse 80, 4056 Basel, Switzerland}
 \address{$^{2}$
 Fritz-Haber-Institut der Max-Planck-Gesellschaft, Faradayweg 4-6, 14195 Berlin, Germany
}
\address{$^{3}$Labaratorie Aim\'e Cotton, CNRS, Universit\'e Paris-Sud, 91405 Orsay, France}

 \address{$^{4}$Institute of Atomic and Molecular Physics, Jilin University, 2699 Qianjin Avenue, Changchun city, 130012, China}
\ead{$^{*}$dongdongzhang@jlu.edu.cn}
\footnote{Present address: European Patent Office, Patentlaan 2, 2288 EE Rijswijk, The Netherlands}
\vspace{10pt}
\begin{indented}
\item[] \today
\end{indented}
\begin{abstract}
The concept of a novel traveling wave Zeeman deccelerator based on a double-helix wire geometry capable of decelerating paramagnetic species with high efficiency is presented. A moving magnetic trap is created by running time-dependent currents through the decelerator coils. Paramagnetic species in low-field-seeking Zeeman states are confined in the moving traps and transported to the end of the decelerator with programmable velocities. Here, we present the theoretical foundations underlying the working principle of the traveling-trap decelerator. 
Using trajectory simulations, we characterise the performance of the new device and explore the conditions for phase-space stability of the transported molecules.

\end{abstract}

\maketitle

\section{Introduction}
In recent years, significant efforts have been invested into developing methods for the production of molecules at cold ($<$1~K) and ultra-cold translational temperatures($<$1~mK)~\cite{carr09a,balakrishnan16a,deMille17a}.
Besides the development of direct laser cooling~\cite{tarbutt18a}, different techniques have emerged for cooling molecules based on their interactions with electric or magnetic fields~\cite{meerakker12a}. 
These developments have been motivated by prospects of studying molecular collisions and chemical reactions at low and precisely controllable collision energies~\cite{narevicius12a,balakrishnan16a,jin12a,Bohn17a}, of precision spectroscopic measurements for testing fundamental physical concepts~\cite{safronova18a,chupp19a}, and of new approaches to quantum-information processing and quantum simulation~\cite{krems09a,georgescu14a,mcardle20a}. Methods based on the deceleration of supersonic molecular beams are particularly well suited for collision experiments since the final longitudinal velocity of the sample can be tuned over a wide range with narrow velocity spreads~\cite{narevicius12a,Kirste12a,vogel14a,Vogels15a,Akerman15a,gao18a,vogels18a,segev19a}. In this context, the Zeeman deceleration method relies on the state-dependent interaction of neutral paramagnetic atoms or molecules with time-dependent inhomogeneous magnetic fields~\cite{vanhaecke07a,Narevicius07a,hogan07a,hogan08a,wiederkehr11a,wiederkehr12a,trimeche11a,Lavert_Ofir11a,motsch14a,liu15a,cremers17a,akerman17a,semeria18a,cremers18a,mcard18a,plomp19a,cremers19a} and is thus suitable for open-shell systems such as molecular radicals or metastable atoms and molecules~\cite{hogan11a,narevicius12a,meerakker12a}.

Here, we present a novel traveling-wave Zeeman decelerator recently developed in our laboratory. The new decelerator operates with traveling magnetic traps containing molecules which are adiabatically slowed down, which is conceptually similar to the traveling-wave Stark decelerator~\cite{meek08a,meek09a,meek09b,osterwalder10a,meek11a,bulleid12a} but works for paramagnetic atoms and molecules. The present paper focuses on the theory and numerical analysis of the operational principle of the device. The experimental implementation and its demonstration are described in the accompanying article~\cite{damjanovic21b}. As areas of application of the new decelerator we envisage cold-collision experiments, trap loading for further cooling and studies of the chemistry of trapped particles at very low temperatures.

\section{Formation of a traveling magnetic trap}



\begin{figure}[t]
    \includegraphics[width=\textwidth]{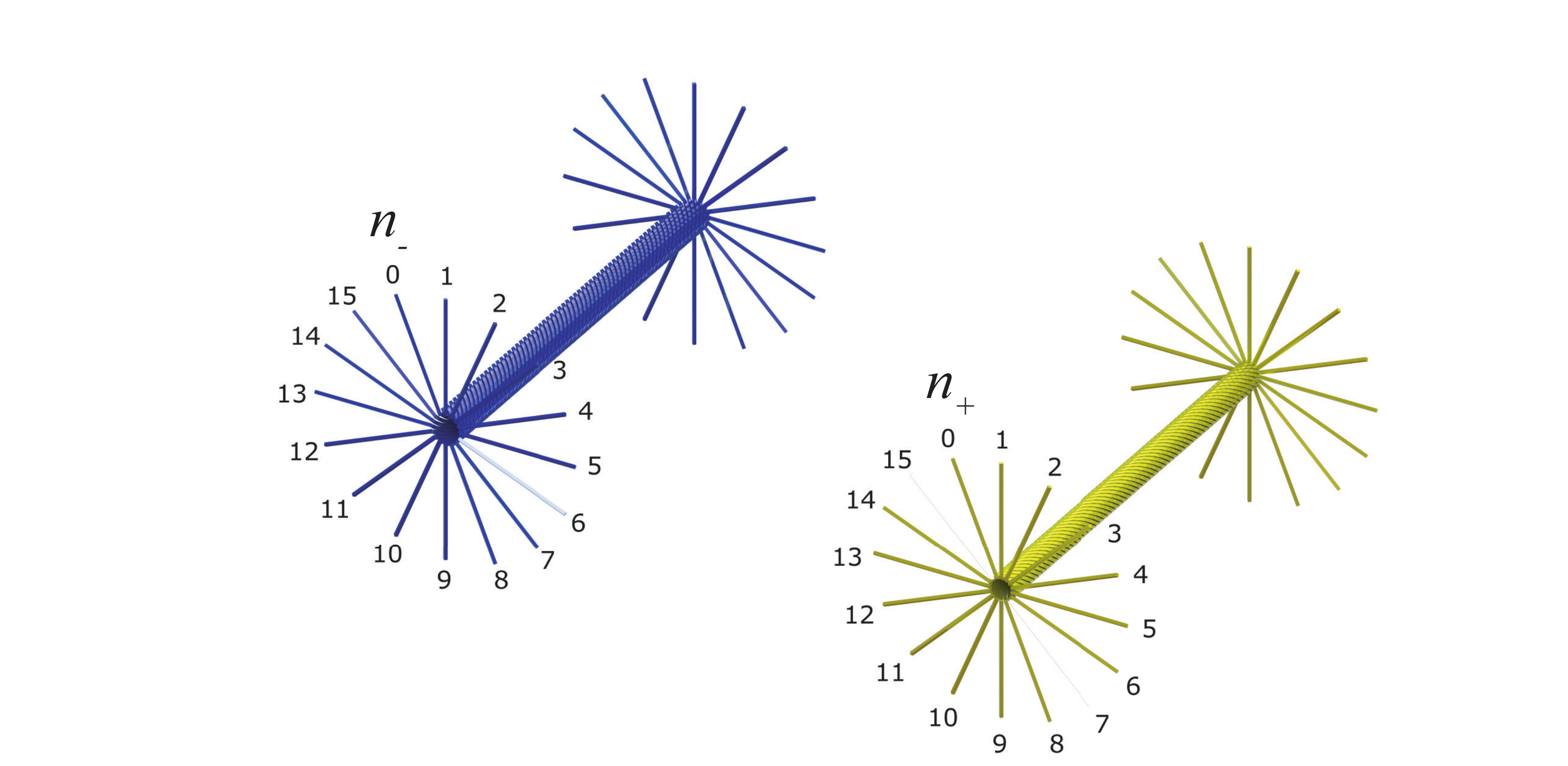}
    \centering
    \caption{Schematic of a decelerator module consisting of two stacked layers of 16 wires wound in right- and left-handed helices. The two layers are depicted separately in blue and yellow for clarity. $n_{-} (n_{+}) \in \{0,..,15\}$ labels the individual wires in the outer, left-handed (inner, right-handed) helix layer.}
    \label{fig:wire}
\end{figure}

\subsection{Longitudinal dynamics of the moving trap}
\label{longitrap}
\begin{figure}[t]
    \includegraphics[width=0.8\textwidth]{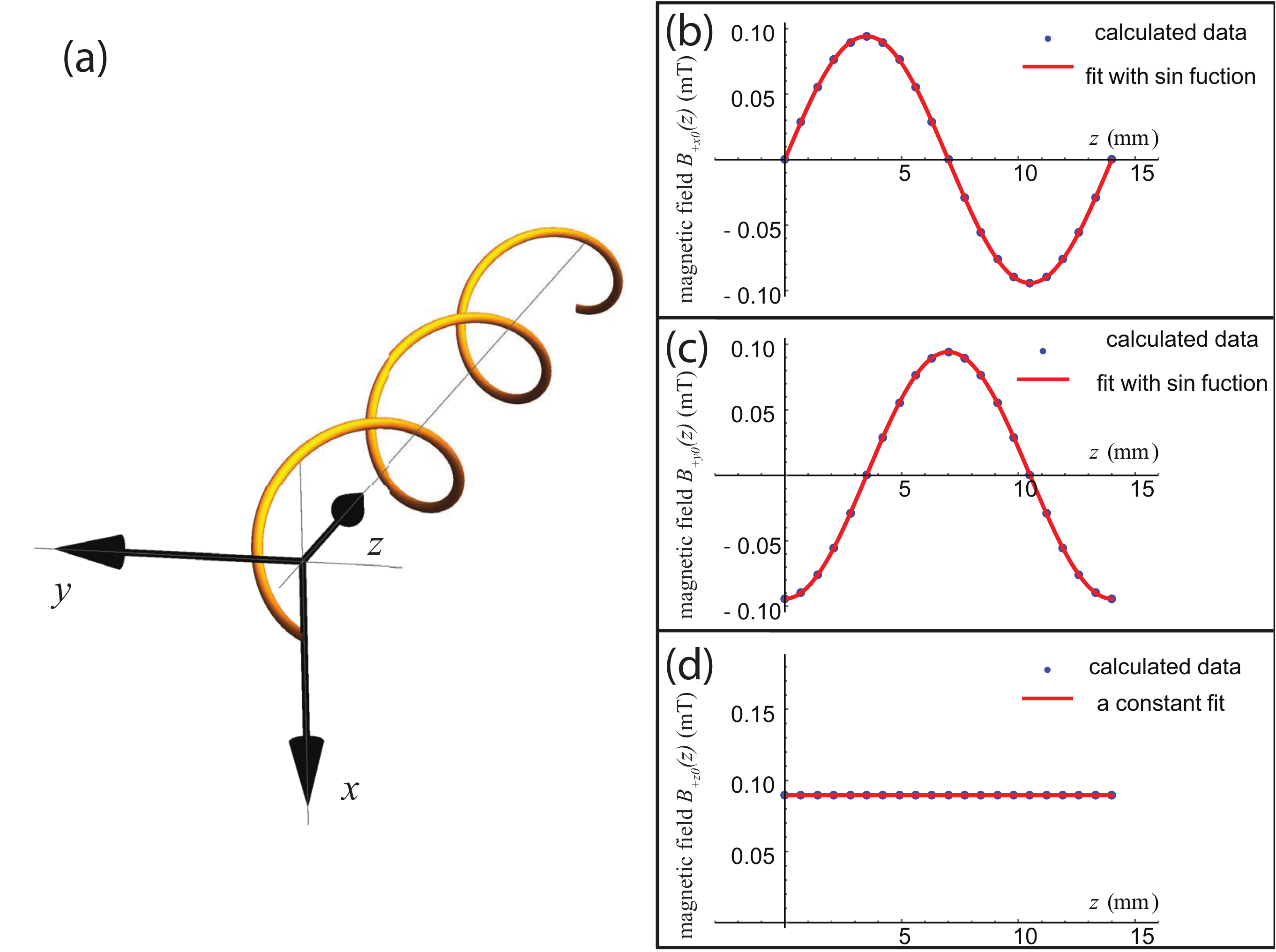}
    \centering
    \caption{(a) Schematic of a single helical wire in a decelerator module. (b),(c) and (d) Magnetic field components $B_{+x0}(z)$, $B_{+y0}(z)$ and $B_{+z0}(z)$ along the central coil axis ($z, x=y=0$) generated from a single wire. Data points represent the results of numerical calculations based on the Biot-Savart law (see text). The lines are a fit of the data to~\Eref{B0p} in the text.}
    \label{fig:magneticfield-singlewire}
\end{figure}
The decelerator features a modular design~\cite{damjanovic21b} with each module consisting of 32 copper wires wound around a cylinder in two layers where the inner (outer) layer consists of 16 right-handed (left-handed) helices, see~\Fref{fig:wire}. The wires are labelled as indicated in the figure. 
By supplying time-varying currents to the double-helix coil geometry, a traveling magnetic wave along the coil axis is formed. To illustrate the principle, we start by analyzing the magnetic fields generated by a single helix within a deceleration module. To a good approximation, the magnetic field on the axis of a coil generated by a current of 1~A supplied to a right-handed wire helix can be expressed as
\begin{eqnarray}
\label{B0p}
    \eqalign{B _{+x0}(z)& = a_{+} \sin(kz), \cr
    B _{+y0}(z)& =  a_{+} \cos(kz),\cr
    B _{+z0}(z)& = a_{+z}},
\end{eqnarray}
and to a left-handed helical coil as
\begin{eqnarray}
\label{B0n}
    \eqalign{B _{-x0}(z)& = -a_{-} \sin(kz), \cr
    B _{-y0}(z)& = a_{-} \cos(kz), \cr
    B _{-z0}(z)& = a_{-z}}.
\end{eqnarray}
The $x_i = x,y,z$ components of the magnetic-field vector generated by a single helical wire, $B_{+/-x_in}$, are shown in~\Fref{fig:magneticfield-singlewire}. The subscript $n = 0,...,15$, denotes the wire index and $+\ (-)$ stands for a right-(left-)hand orientation of the wires. $k = 2\pi /\lambda$ is the wave number of the traveling wave, where $\lambda$ = 14~mm is the periodicity of the helices in the present implementation. The parameters $a_{+/-}$ represent the amplitude of the magnetic field at a probe current of 1~A. Their values are derived in the supplementary material, where also the validity of~\Eref{B0p} and~\Eref{B0n} is demonstrated.\\
It is straightforward to obtain equivalent expressions for the magnetic fields produced by the other wires with a rotational transformation by an angle $n\Delta$ where $\Delta$ = 2$\pi$/16 due to 16 wires being distributed evenly around a cylinder. For the right- and left-handed helices, the components of the magnetic field are accordingly given by
\begin{eqnarray}
\label{Bnn}
    \eqalign {B _{+xn}(z)& = a_{+} \sin(kz + n\Delta),\cr
    B _{+yn}(z)& = a_{+} \cos(kz + n\Delta),\cr
    B _{+zn}(z)& = a_{+z}},
\end{eqnarray}
\begin{eqnarray}
\label{Bpn}
    \eqalign {B _{-xn}(z)& = -a_{-} \sin(kz + n\Delta),\cr
    B _{-yn}(z)& = a_{-} \cos(kz + n\Delta),\cr
    B _{-zn}(z)& = a_{-z}}.
\end{eqnarray}
Time-dependent currents of the form
\begin{eqnarray}
\label{current}
    \eqalign {c _{+n}(t)& = c_{+} \sin(\phi_{+}(t) + n\Delta),\cr
    c _{-n}(t)& = c_{-} \sin(\phi_{-}(t) + n\Delta)}
    \label{eq:tdpcurr}
\end{eqnarray}
are applied to the $n$-th wire, where $\phi_{+/-}(t)$ is the time-dependent phase of the currents supplied to the right- (left-) handed layer. The additional phase $n\Delta$ serves as a compensation to the geometrical arrangement of the wires in order to minimise higher-order harmonics in the synthesised magnetic-field profiles. $c_{+/-}$ are the current amplitudes for left-(+) and right-(-) handed wires, respectively, and can be tuned individually. 
The magnetic-field components generated by the entire double-layer coil geometry are expressed as
\begin{equation}
\begin{split}
    \label{Bx}
       \eqalign{B_{x}(z,t)  &=  \sum\limits_{n=0}^{15} \big\{  c_{+}  a_{+}  \sin\big[\phi_{+}(t) + n\Delta\big]  \sin(kz + n\Delta)\\
  &- c_{-}  \sin\big[\phi_{-}(t) + n\Delta\big]  a_{-}  \sin(kz + n\Delta) \big\}, }
\end{split}
\end{equation}\\
\begin{equation}
\begin{split}
    \label{By}
         \eqalign{B_{y}(z,t) & =        \sum\limits_{n=0}^{15} \big\{ c_{+}  \sin\big[\phi_{+}(t) + n\Delta\big]  a_{+}  \cos(kz + n\Delta)\\
   &+ c_{-}  \sin\big[\phi_{-}(t) + n\Delta\big]  a_{-}  \cos(kz + n\Delta) \big\},}
   \end{split}
\end{equation}\\
\begin{equation}
\begin{split}
   \label{Bz}
       \eqalign{B_{z}(z,t)& =   \sum\limits_{n=0}^{15} \big\{  c_{+}  \sin\big[\phi_{+}(t) + n\Delta\big]  a_{+z}\\
  &+ c_{-}  \sin\big[\phi_{-}(t) + n\Delta\big]  a_{-z} \big\} = 0.}
  \end{split}
\end{equation}
By introducing new parameters $A_\mathrm{avg}$ and $\Delta A$ which are the sum and the difference, respectively, of the geometry-weighted current amplitudes applied to the two layers, $c_{+}a_{+}$ and $c_{-}a_{-}$ can be expressed as
\begin{eqnarray}
    \label{amplitude}
    \eqalign {c_{+}a_{+} = A_\mathrm{avg}+\frac{\Delta A}{2},\cr
    c_{-}a_{-} = A_\mathrm{avg}-\frac{\Delta A}{2}}.
\end{eqnarray}
The phases $\phi _{+}(t)$, $\phi _{-}(t)$ can be written as
\begin{eqnarray}
	\label{phase}
    \eqalign {\phi_{-}(t) = \phi_\mathrm{avg}(t) + \frac{\Delta\phi(t)}{2},\cr
    \phi_{+}(t) = \phi_\mathrm{avg}(t) - \frac{\Delta\phi(t)}{2}}.
\end{eqnarray}
The ratio $c_+a_+/c_-a_-$ depends on the radius and the current amplitude of each layer and can be tuned in a way such that $\Delta A \approx0$.
Inserting~\Eref{amplitude} and~\Eref{phase} into~\Eref{Bx} and~\Eref{By} and assuming $\Delta A \approx0$ yields
\begin{eqnarray}
    \label{bsx}
    B_{x}(z,t) = NA_\mathrm{avg} \sin(kz - \frac{\Delta \phi(t)}{2})  \sin\left(\phi_\mathrm{avg}(t)\right),
\end{eqnarray}
\begin{eqnarray}
    \label{bsy}
    B_{y}(z,t) = NA_\mathrm{avg} \sin\left(kz - \frac{\Delta \phi(t)}{2}\right)  \cos(\phi_\mathrm{avg}(t)),
\end{eqnarray}
\begin{eqnarray}
    B_{z}(z,t) = 0,
\end{eqnarray}
where $N = 16$. Thus, the magnitude of the magnetic field on the central axis of the full double-helix assembly reduces to the form of a traveling wave:
\begin{equation}
    |\mathbf B(z,t)| = \sqrt{B_x^2+B_y^2+B_z^2} = NA_\mathrm{avg}\left|\sin\left(kz-\frac{\Delta \phi(t)}{2}\right)\right|.
    \label{eqn:tw1}
\end{equation}
\begin{figure}[t]
    \includegraphics[width=0.6\textwidth]{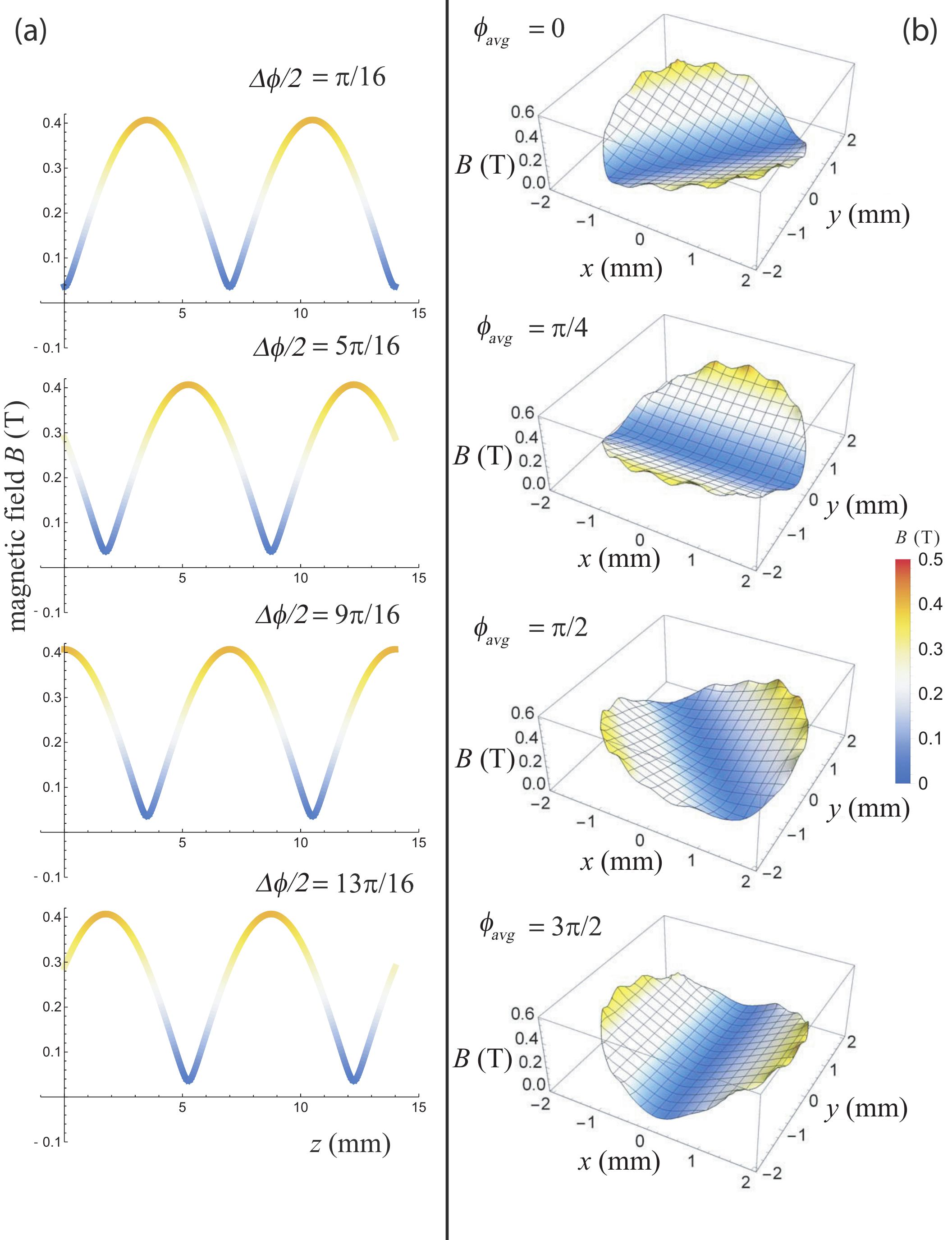}
    \centering
    \caption{Trap dynamics as a function of the phases $\Delta \phi /2$ (a) and $\phi_\mathrm{avg}$ (b) controlled by applying appropriate time dependent currents to the two layers of decelerator coils. The current amplitude used in the calculation is 300~A which is the typical value chosen in the experiments.}
    \label{fig:trapcontrol}
\end{figure}
As can be seen in~\Eref{bsx} and~\Eref{eqn:tw1}), the dynamics of the magnetic wave is controlled by the two time-dependent phases $\frac{\Delta \phi(t)}{2}$ and $\phi_\mathrm{avg}(t)$. The phase $\Delta \phi(t)$ serves to control the position of the wave along the coil axis. By applying a time dependence of the form
\begin{equation}
    \frac{\Delta \phi(t)}{2} = At^2+Bt+C,
\end{equation}
the traveling magnetic wave can be programmed to decelerate, accelerate or 
propagate with constant velocity along the $z$ coordinate. The parameters $A, B$ and $C$ are given by
\begin{eqnarray}
A = \frac{k}{2}\frac{v_\mathrm F^2 - v_\mathrm I^2}{2L}, \cr
B = kv_\mathrm I, \cr
C = 0.
\end{eqnarray}
$A$ determines the deceleration (acceleration) and $B$ encodes the initial velocity $v_\mathrm{I}$ of the traveling wave. The constant phase shift $C$ can be set to zero for practical purposes. $v_\mathrm F$ stands for the final velocity of the traveling wave and $L$ is the length of the decelerator. The on-axis trap dynamics as a function of time-dependent phase $\frac{\Delta \phi (t)}{2}$ is illustrated in~\Fref{fig:trapcontrol} (a) where the propagation of the magnetic wave is illustrated by plots of the magnetic field for four values of the phase $\frac{\Delta \phi}{2} = \frac{\pi}{16}, \frac{5\pi}{16},\frac{9\pi}{16}$ and $\frac{13\pi}{16}$.

So far, only the magnetic-wave dynamics on the central coil axis was discussed. The transverse, i.e., perpendicular to the central axis, components of the magnetic field were calculated numerically as explained in the supplementary material. The transverse dynamics of the magnetic wave is controlled by the time-dependent phase $\phi_\mathrm{avg}(t)$. This is illustrated in~\Fref{fig:trapcontrol} (b) where the calculated magnetic field $|\mathbf B(x,y,0,t)|$ is plotted for four different values $\Delta \phi_{avg} = 0, \frac{\pi}{4}, \frac{\pi}{2}$ and $\frac{3\pi}{2}$. 
As can be seen in~\Fref{fig:trapcontrol} (b), the magnetic field in the transverse direction exhibits a deep minimum along one direction capable of confining magnetically low-field-seeking species. In the perpendicular direction, the minimum is very shallow so that particles can escape along this axis. Changing the phase $\Delta \phi_\mathrm{avg}$ varies the orientation of the trap in the transverse direction. This effect can be understood as follows: assume a stationary on-axis magnetic wave, i.e., \big($\frac{\Delta \phi}{2}(t) = 0$\big), then the magnitude of the magnetic field in the transverse direction (at $z=0$) generated by the $n$-th right- and left-hand-oriented helix is given by
\begin{eqnarray}
\begin{split}
      \mathbf B_n(x,y,0,t) =& c_+\mathbf B_{+n}(x,y,0)\sin(\Delta\phi_\mathrm{avg}(t)+n\Delta)\\ 
    &+c_-\mathbf B_{-n}(x,y,0)\sin(\Delta\phi_\mathrm{avg}(t)+n\Delta),
    \label{eq:tr_bf}
    \end{split}
\end{eqnarray}
where $\mathbf B_{+/-n}(x,y,0)$ is the magnetic field in the $xy$ plane ($z=0$) generated by the probe current supplied to the $n$-th helix in right (left) layer. The right- and left-handed $n$-th helices are maximally contributing to the total magnetic field when $\Delta \phi_\mathrm{avg}+n\Delta=\frac{k\pi}{2}, k = 0,1,2,...$ If the phase $\Delta \phi_\mathrm{avg}(t)$ changes with time as $\Delta\phi_\mathrm{avg} = \omega_\mathrm{avg}t$ and an angular frequency $\omega_\mathrm{avg}$, the time at which the $n$-th helices are maximally contributing is $t_n = \frac{k\pi-n\Delta}{2\omega_\mathrm{avg}}$. At the time $t_{n+1} = t_n+\frac{\Delta}{2\omega_\mathrm{avg}}$, the maximal contribution comes from the ($n+1$)-th helices and the transverse field was rotated by an angle $\Delta$. In this way, the transverse field is rotated with time at the angular frequency $\omega_\mathrm{avg}$. The two phases $\frac{\Delta \phi}{2}(t)$ and $\phi_\mathrm{avg}(t)$ can be chosen independently from each other, leading to a decoupling of the longitudinal and transverse motions of the trap. This feature prevents trap losses due to the motional coupling which are characteristic of the conventional Stark and Zeeman decelerators \cite{meerakker05c,wiederkehr10a}. Experimentally, both phases are independently controllable through the time-dependent phases of the supplied currents $\phi_+(t)$ and $\phi_-(t)$, ~\Eref{phase}.
\begin{figure}[h]
    \includegraphics[width=0.6\textwidth]{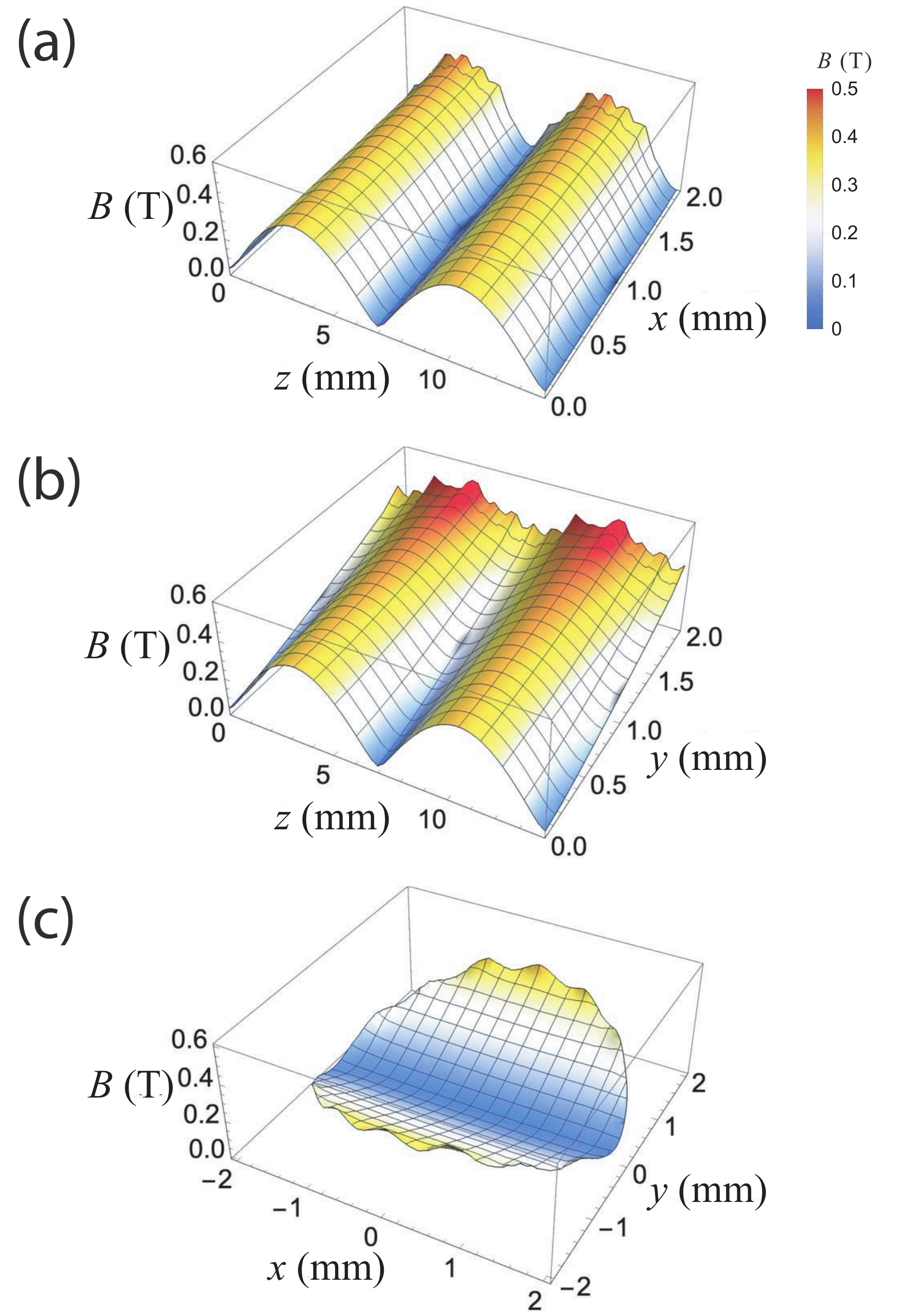}
    \centering
    \caption{Magnetic field generated by applying currents of 300~A to both layers of the decelerator coils in the (a) $zx$, (b) $zy$ and (c) $xy$ planes.
    }
    \label{fig:modbfield}
\end{figure}
As an example, numerically calculated magnetic fields in $zx$, $zy$ and $xy$ planes at $t$ = 0 generated by time-varying currents as given by~\Eref{eq:tdpcurr} and a current amplitude $I$ = 300~A are shown in~\Fref{fig:modbfield} (a)-(c). The magnetic field exhibits confining characteristics along the $z$ and $y$ coordinates, but only weak trapping along the $x$ coordinate as discussed above (see also~\Fref{fig:anharm}). 

\subsection{Transverse stability}
\label{sec:transstab}

\begin{figure}[h]
    \includegraphics[width=\textwidth]{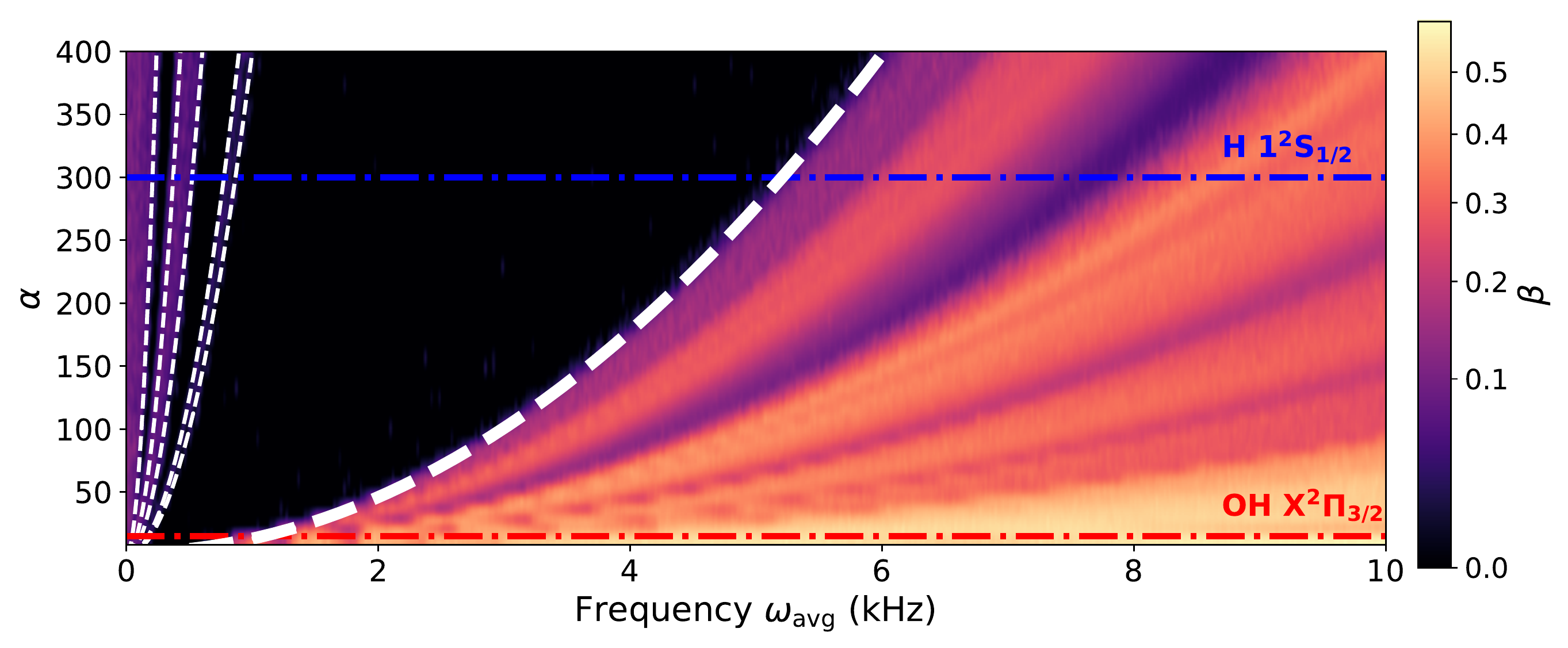}
    \caption{Stability diagram of a rotating magnetic trap plotted plotted as a function of the rotational frequency $\omega_\mathrm{avg}$ and the parameter $\alpha$ (see text). Dark and bright colors represent regions of low and high stability, respectively. Black areas constitute regions in which no stable trajectories were found. Regions of stable and unstable behaviour are separated by white dashed lines. The threshold frequency (thick white dashed line) required for uniformly stable trajectories is proportional to $\propto \sqrt{\alpha}$.}
    \label{fig:stab}
\end{figure}

\begin{figure}[h]
    \centering
    \includegraphics[width=0.8\textwidth]{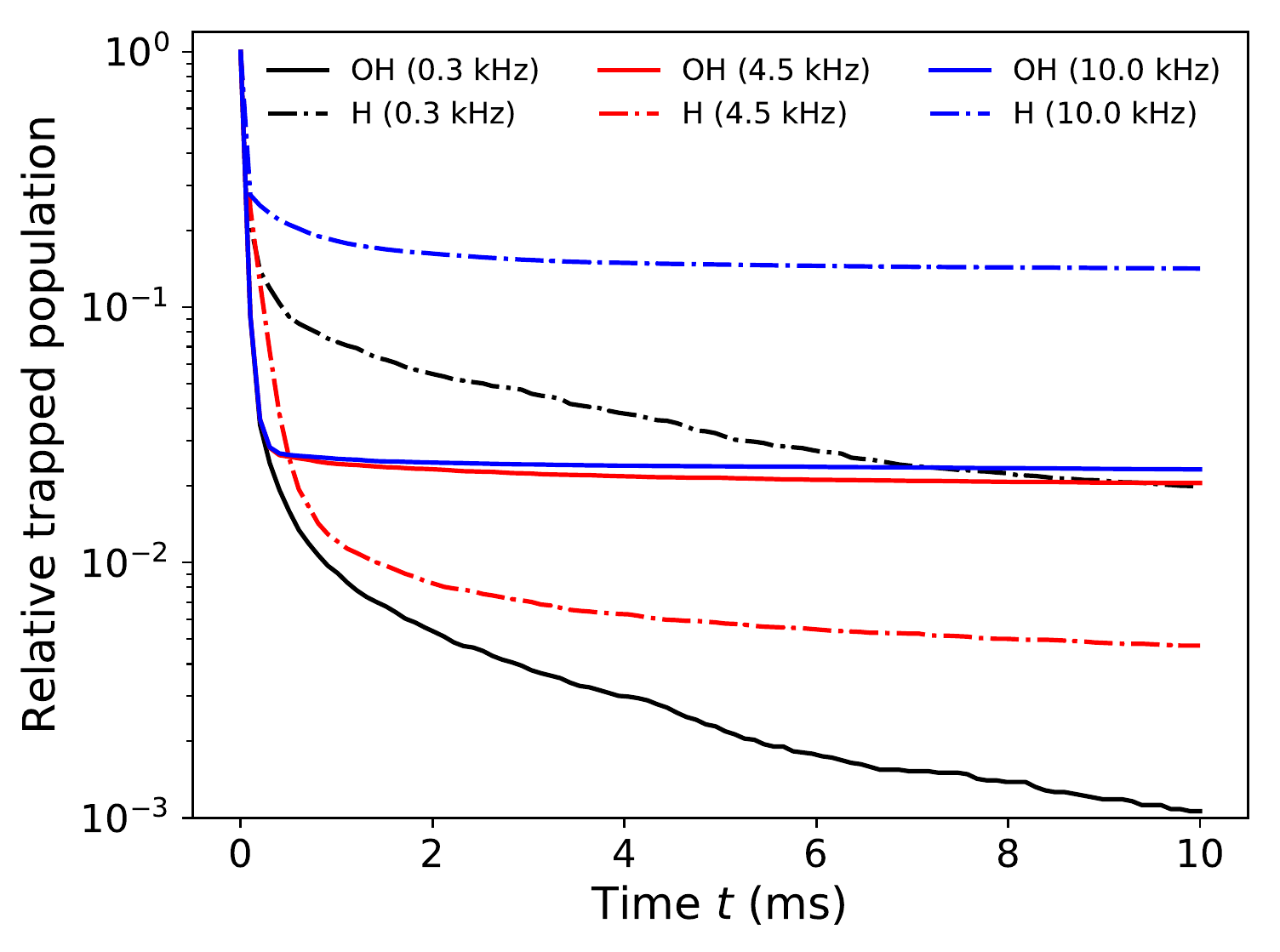}
    \caption{Relative population of OH molecules and H atoms inside a rotating trap at different instances of time for three different values of the rotational frequency $\omega_\mathrm{avg}$ = 0.3 kHz, 4.5 kHz and 10 kHz.}
    \label{fig:rtp1}
\end{figure}

In this section, the transverse stability of particles confined in a rotating trap is discussed. The equations of motion along the $x$ and $y$ directions are given by the coupled differential equations
\begin{equation}
    \ddot x(t) = -\alpha \frac{\mu_{\mathrm B}}{m_\mathrm{u}}\frac{\partial}{\partial x}|\mathbf{B}_0\big(x(t),y(t),0,t\big)|,
    \label{eq:eq_stab1}
\end{equation}
\begin{equation}
    \ddot y(t) = -\alpha \frac{\mu_{\mathrm B}}{m_\mathrm{u}}\frac{\partial}{\partial y}|\mathbf{B}_0\big(x(t),y(t),0,t\big)|.
    \label{eq:eq_stab2}
\end{equation}
Here, $\mathbf{B}_0\big(x(t),y(t),z(t),t\big)$ is the magnetic field generated by a probe current of 1~A supplied to the coil assembly. The magnetic field generated by an arbitrary current is obtained by scaling the $\mathbf{B}_0\big(x(t),y(t),z(t),t\big)$ by a factor $c=c_+=c_-$, $\mathbf{B}\big(x(t),y(t),z(t),t\big) = c\mathbf{B}_0\big(x(t),y(t),z(t),t\big)$. $\alpha $ is a dimensionless parameter given by $\alpha = dc$, where $d = \frac{\mu_\mathbf{eff}}{M}$ is the magnetic-dipole-moment-to-mass ratio, $\mu_\mathrm{eff}$ is the effective magnetic dipole moment and $M$ is the mass of the particle (in atomic units). $\mu_\mathrm{B}$ is the Bohr magneton and $m_\mathrm{u}$ is the atomic mass constant. 
\begin{figure}[h]
    \centering
    \includegraphics[width=\textwidth]{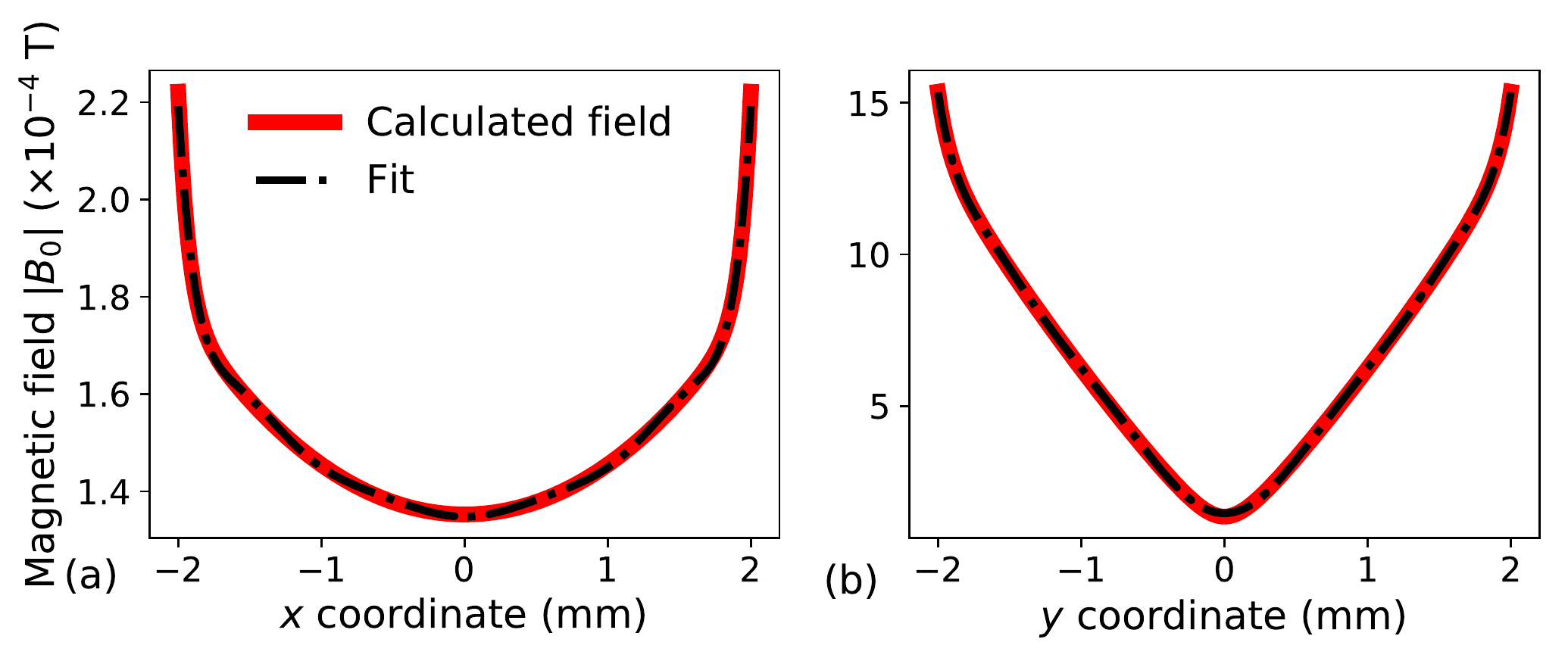}
    \caption{Magnetic field $|\mathbf B_0(t)|$ along (a) $x$ and (b) $y$ coordinate. }
    \label{fig:anharm}
\end{figure}
To discuss the dynamics of the particles in the rotating trap, the magnetic field $|\mathbf B_0|$ along the $x$ and $y$ coordinates at $t=0$ are plotted in~\Fref{fig:anharm}. The calculated magnetic field is shown by the red lines. A polynomial fit using the function $f_\mathrm{fit} = a_0+a_2x_i^2+a_4x_i^4+a_6x_i^6+a_8x_i^8+a_{10}x_i^{10}$, $x_i=x,y$ is shown by the black dash-dotted lines. By visual inspection of~\Fref{fig:anharm} and from the fitted parameters listed in~\Tref{tab:tab1}, it becomes apparent that the magnetic field along both directions exhibits a high degree of anharmonicity with 4-th- and 6-th-order terms still being significant. 
\begin{table}[htb]
\caption{Polynomial fit coefficients for the magnetic field $|\mathbf B_0|$ shown in \Fref{fig:anharm} along the $x$ and $y$ coordinates. }
\centering\label{t:observed_psrs}
\begin{tabular}{lll}\noalign{\smallskip} \hline \hline \noalign{\smallskip}
  & $x$ coordinate &  $y$ coordinate\\\hline
$a_0$           & $1.464(8)\times 10^{-4}$              & $1.348(2)\times 10^{-4}$ \\
$a_2$           & $8.39(6)\times 10^{-4}$               & $1.78(8)\times 10^{-5}$ \\
$a_4$           & $-6.0(2)\times 10^{-4}$               & $-2.1(2)\times 10^{-5}$ \\
$a_6$           & $ 3.18(8)\times 10^{-4}$               & $1.9(1)\times 10^{-5}$\\
$a_8$           & $ -8.4(2)\times 10^{-5}$              & $-7.1(3)\times 10^{-6}$ \\
 $a_{10}$          & $8.6(2)\times 10^{-6}$            & $9.1(3)\times 10^{-7}$ \\
\noalign{\smallskip} \hline \noalign{\smallskip}\end{tabular}
\label{tab:tab1}
\end{table}
As a result of the anharmonicity, analytical solutions to the coupled differential equations~\Eref{eq:eq_stab1} and~\Eref{eq:eq_stab2} do not exist. Therefore, the stability of the solutions as a function of the parameter $\alpha$ and the rotational frequency $\omega_\mathrm{avg}$ was examined numerically. Solutions were propagated in time up to $t=50\ \mathrm{ms}$ using an adaptive 4-th order Runge-Kutta algorithm. The positions of the particles at the end of the propagation were examined. If the position was found to lie within the trapping region, the trajectory was considered as stable, and otherwise as unstable. For given parameters $\alpha$ and $\omega_\mathrm{avg}$, the equations were solved numerically and the parameter $\beta = \tfrac{\big\langle\big|\mathbf{B}_{0,\alpha}^\mathrm{max}\big(x(t),y(t),0,t\big)\big|\big\rangle}{\big\langle\big|\mathbf{B}_0^\mathrm{max}\big(x(t),y(t),0,t\big)\big|\big\rangle}$ was calculated. The pointed brackets denote a time average. $\beta$ represents the ratio of the maximum of the time-averaged magnetic field $\mathbf B_0$ for which the trajectory of a particle is stable and the time-averaged maximum of the magnetic field $\mathbf B_0$, or in other words, the ratio of the effective trapping potential and the maximum trapping potential in the limit $\omega_\mathrm{avg}\rightarrow\infty$ for a given particle. \\
The resulting stability diagram of the trap is shown in~\Fref{fig:stab}. Solutions of the equations of motion were explored for a frequency range $\omega_\mathrm{avg} = 2\pi\times(0-10)\ \mathrm{kHz}$, which corresponds to the operational range of the experimental decelerator. The dependence of the parameter $\beta$ on the parameters $\alpha$ and $\omega_\mathrm{avg}$ is illustrated by a color map in~\Fref{fig:stab}. Regions of unstable trajectories appear in black. Seven distinct regions were found within the parameter space studied, four regions with stable and three with unstable trajectories. White dashed lines delineate the borders between stable and unstable regions, and were produced by fitting a harmonic function $k\omega_\mathrm{avg}^2$ to the data where $k$ is a fitting parameter. \\
The highest stability (maximum $\beta$) is achieved for low values of $\alpha$ where trajectories are stable across the majority of the range of frequencies. With increasing $\alpha$, the threshold frequency indicated by the thick white dashed line above which there are no regions of unstable behaviour increases and is proportional to $\sqrt{\alpha}$. The existence of regions of stable trajectories both at low and high rotational frequencies allows, in principle, for a selective confinement of multiple molecular species in the deceleration process. This property is illustrated by the example of the simultaneous trapping of H atoms in the $1^2S_{1/2}(m_s = 1/2)$ state (blue dash-dotted line) and OH molecules in the $X^2\Pi_{3/2}(\nu=0,m_J = 3/2)$ state (red dash-dotted line). At low rotational frequencies (e.g. $\omega_\mathrm{avg} = 2\pi\times0.3$ kHz, second region of stability) H atoms have stable trajectories while OH molecules have unstable trajectories and are thus ejected from the trap. Conversely, if the rotational frequency is in an intermediate range $\omega_\mathrm{avg} \lessapprox 2\pi\times5$~kHz, only trajectories of the OH molecules are stable. In the frequency range $\omega_\mathrm{avg} \gtrapprox 2\pi\times5$~kHz, trajectories of both species are stable and thus both species are trapped efficiently. \\
This effect is illustrated by numerically solving~\Eref{eq:eq_stab1} and~\Eref{eq:eq_stab2} for $10^5$ OH molecules and H atoms initialised in the trap with a uniform spatial distribution and a velocity distribution of 20~m/s (FWHM) for three rotational frequencies $\omega_\mathbf{avg}$ = 0.3 kHz, 4.5 kHz and 10 kHz. The relative population of each species inside the trap is extracted at different trapping times. The results are shown in~\Fref{fig:rtp1}. The simulations confirm the enhanced trapping of H atoms for $\omega_\mathrm{avg} =0.3 \mathrm{\ kHz}$ (black lines) relative to OH and vice versa for $\omega_\mathrm{avg}$ = 4.5 kHz (red lines). For $\omega_\mathrm{avg}$ = 10 kHz (blue lines), both species show stable confinement with time with OH exhibiting a lower relative trapped population due to the lower trapping potential. These effects are appreciable at the timescales of our experiments and will be the subject of future experimental studies.

\subsection{Numerical trajectory simulations}

In order to model the particle dynamics within the entire decelerator and characterize its deceleration performance, a numerical trajectory simulation code in the Python programming language was developed. In the simulation, $10^5$ OH molecules were initialised and their trajectories through the decelerator were calculated under different operating conditions. 

\begin{figure}
\includegraphics[width=0.6\textwidth]{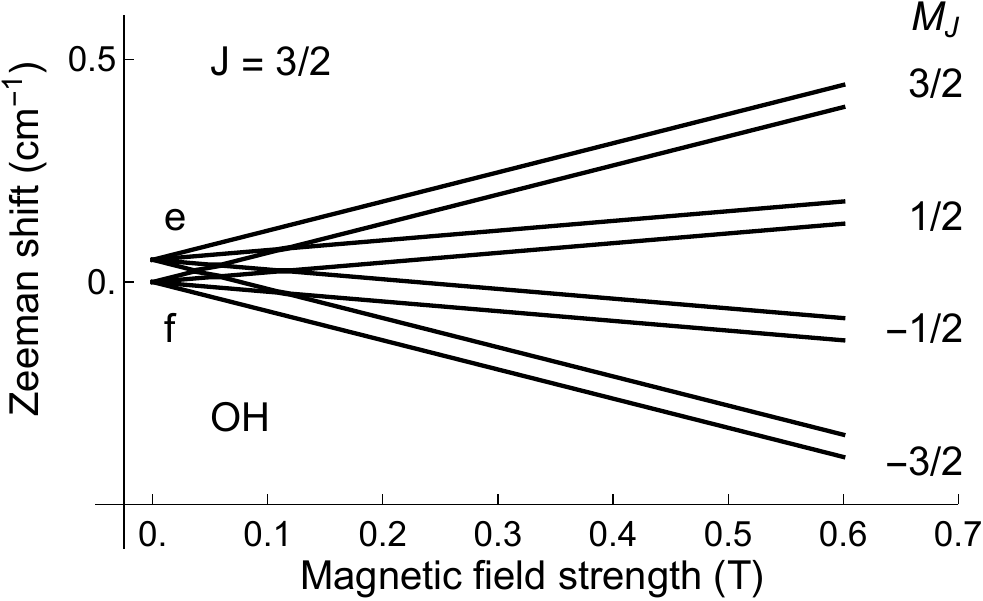}
\centering
\caption{The Zeeman effect of OH in its ground rovibronic state, $X ^{2}\Pi _{3/2}, v=0, J =3/2$, as a function of magnetic field strength $B$. $e$ and $f$ denote the parity components of the $\Lambda$-doublet.}
\label{fig:zeeman}
\end{figure}

The forces $F_i$ acting on each particle $i$ along its trajectory $x_i(t))$ were obtained from its Zeeman energy $W_i$ according to $F_i(x_i(t))=-\nabla W_i(x_i(t))$. The Zeeman energy level structure of OH~\cite{brown03a} in its $X^2\Pi_{3/2}, v=0$ ground electronic and vibrational state in a magnetic field was calculated according to
\begin{equation}
\label{equ:zeemaneffect}
W = \mu _{eff} B = \mu_{B}(\Lambda+g_{s}\Sigma)\frac{M_{J}\Omega_{eff}}{J(J+1)}B
\end{equation} 
where $\Lambda$ and $\Sigma$ are the quantum numbers of the projection of the electron orbital and spin angular momenta on the molecular axis, $g_{s}$ is the electron-$g$ factor ($g_s\approx$2), and $M_J$ and $\Omega_{eff}$ are the quantum numbers of the projection of the total angular momentum $\vec{J}$ (with quantum number $J$) along the external magnetic field and the molecular axis, respectively~\cite{brown03a}. The Zeeman energy level structure for the lowest rotational level, $J=3/2$, is shown in~\Fref{fig:zeeman}. The low-field seeking quantum states with the largest Zeeman energy at a given magnetic field strength which can be decelerated are $X ^{2}\Pi _{3/2},~|\Omega|=3/2,~e~\text{and}~f,~M_{J} = 3/2$. 

\begin{figure}[t]
\includegraphics[width=0.8\textwidth]{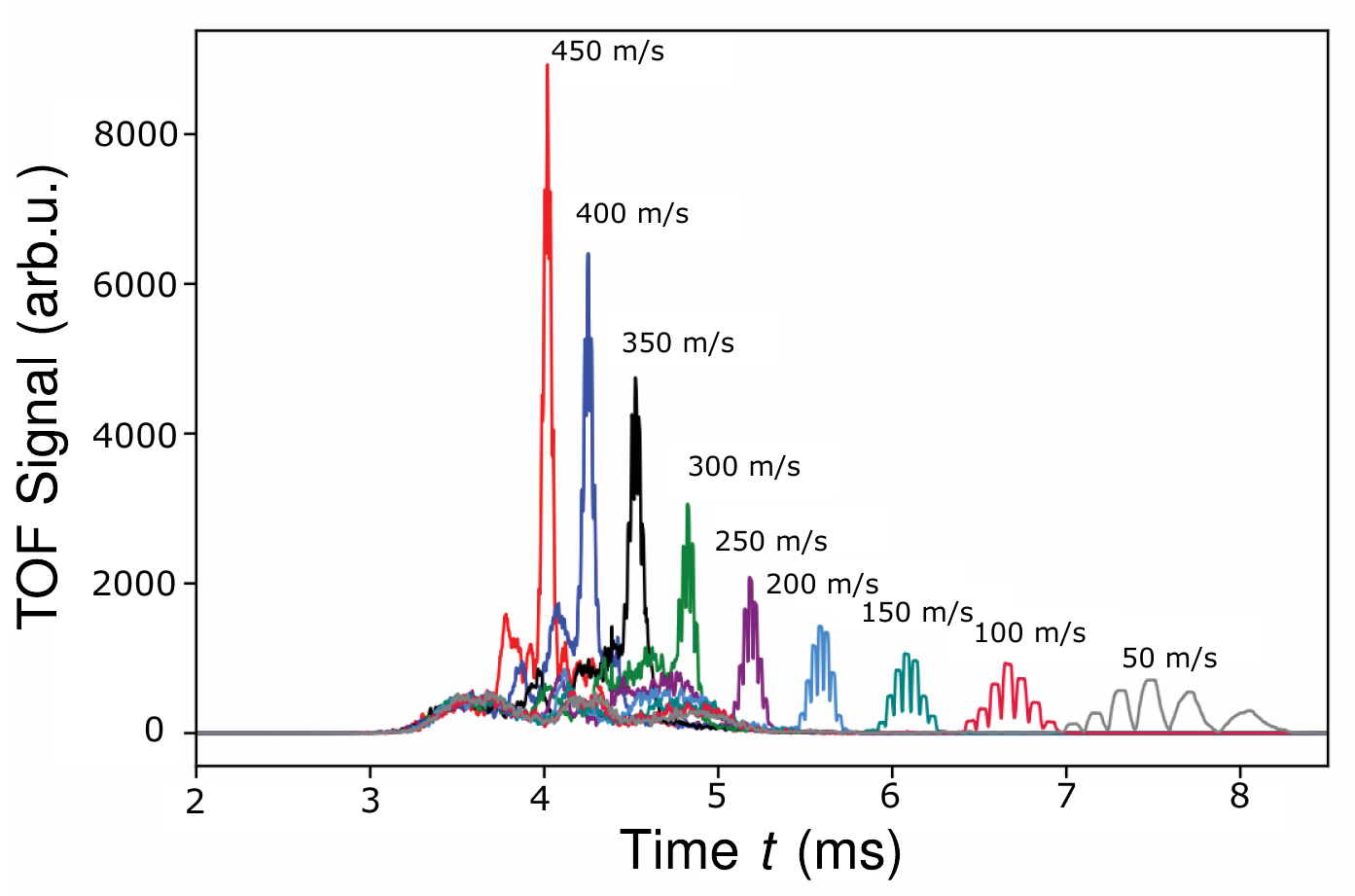}
\centering
\caption{Simulated time-of-flight traces of decelerated OH molecules in the $X^2\Pi_{3/2}, J=3/2, M_J = 3/2$ state. The molecules were decelerated from an initial mean forward velocity of 450~m/s down to 400~m/s, 350~m/s, 300~m/s, 250~m/s, 200~m/s, 150~m/s, 100~m/s and 50~m/s.}
\label{fig:tof_data}
\end{figure}
In the simulations, only molecules in the $M_J = 3/2$ state were considered. The molecules were initially generated within a cylinder with a radius of 2~mm and a length of = 11~mm, which mimics the initial spatial distribution of our experimental molecular beam ~\cite{haas19a}, with a transversal velocity spread of 45~m/s and a longitudinal velocity spread of 90~m/s. The mean forward velocity was initially chosen to be 450~m/s in line with our experiments~\cite{damjanovic21b}. Thus, the particles fill an initial 6-dimensional phase-space volume of $0.5\times 10^8~\mathrm{mm^3(m/s)^3}$. Due to the time-dependence of the currents and complex structure of the wire geometry, the magnetic fields could not be fully calculated prior to a simulation run and were thus calculated on the fly. 
In order to reduce the computation time, the calculations were GPU-accelerated using the NumbaCUDA library in Python~\cite{lam2015numba15a}. In the simulations, the decelerator was assumed to be 1.792~m long (32 deceleration modules~\cite{damjanovic21b}) followed by a 15~mm region of free flight after which the molecules were detected. Exemplary results of time-of-flight profiles of the molecules extracted from the trajectory simulations are displayed in~\Fref{fig:tof_data}. Here, the molecules were decelerated to nine different final velocities in the range 449~m/s - 50~m/s corresponding to decelerations of 0.25~$\mathrm{km/s}^2$ - 55.80~$\mathrm{km/s}^2$. Depending on the final velocity, the total transit times varied from 4.1~ms to 8.2~ms. Owing to the periodic geometry of the traveling wave and depending on the initial longitudinal phase-space volume of molecules, several moving traps can be filled and decelerated. This is evidenced by the substructure visible in the time-of-flight traces in~\Fref{fig:tof_data}, especially at lower final velocities at which neighbouring traps are increasingly separated in time.  
  
\begin{figure}[t]
\includegraphics[width=0.8\textwidth]{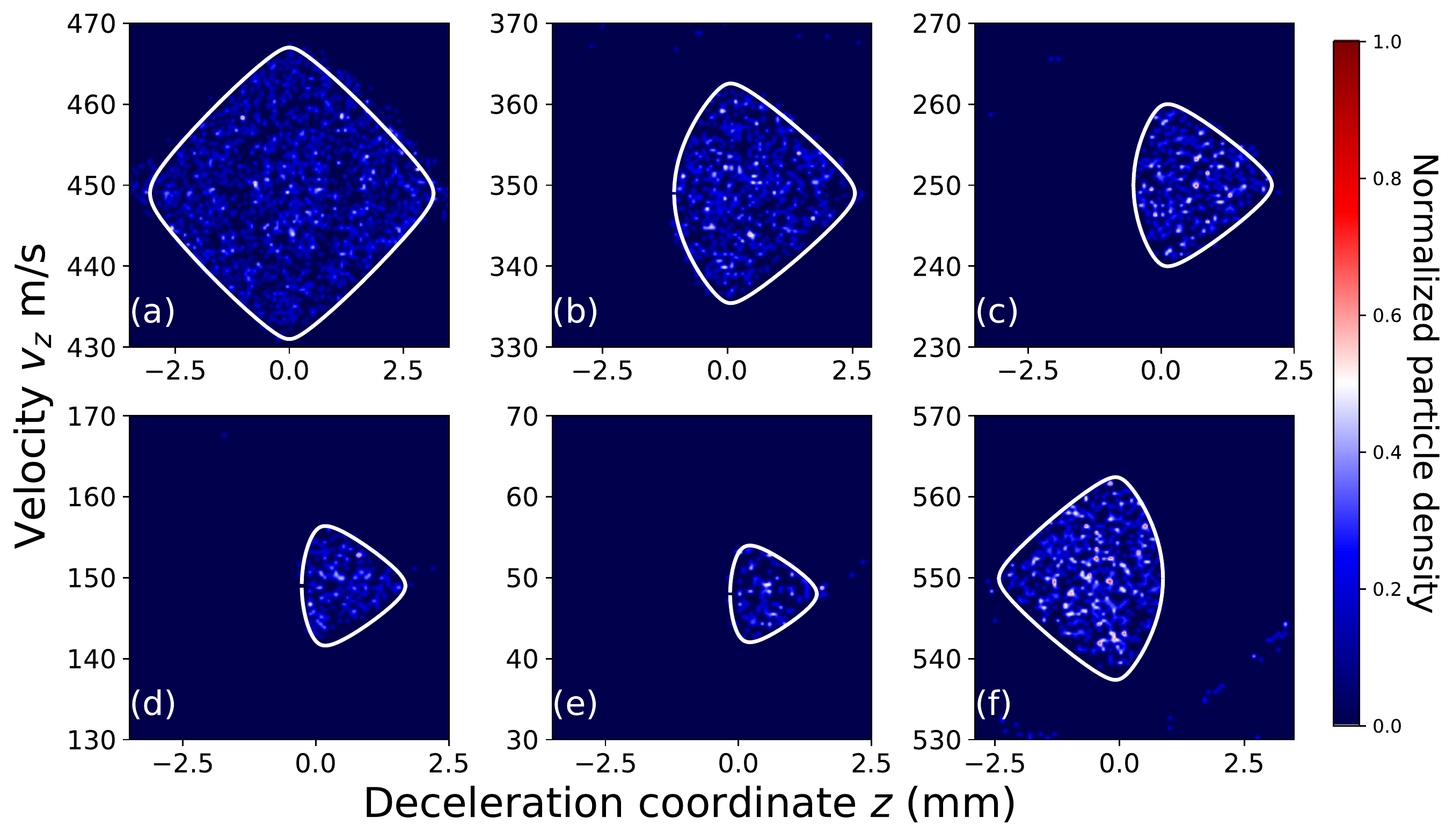}
\centering
\caption{Longitudinal phase-space distributions of molecules obtained from the numerical trajectory simulations at the end of the decelerator for  decelerations (final velocities) of (a) 0.25~km/s$^2$ (449~m/s), (b) 22.5~km/s$^2$ (350~m/s), (c) 39.2~km/s$^2$ (250~m/s), (d) 50.3~km/s$^2$ (150~m/s), (e) 55.8~km/s$^2$ (50~m/s) and (f) -27.6 km/s$^2$ (550~m/s). The white traces delineate calculated phase-stable regions obtained from a 1D model~\cite{wiederkehr10a}.}
\label{fig:psa}
\end{figure}    

\begin{figure}[t]
\includegraphics[width=\textwidth]{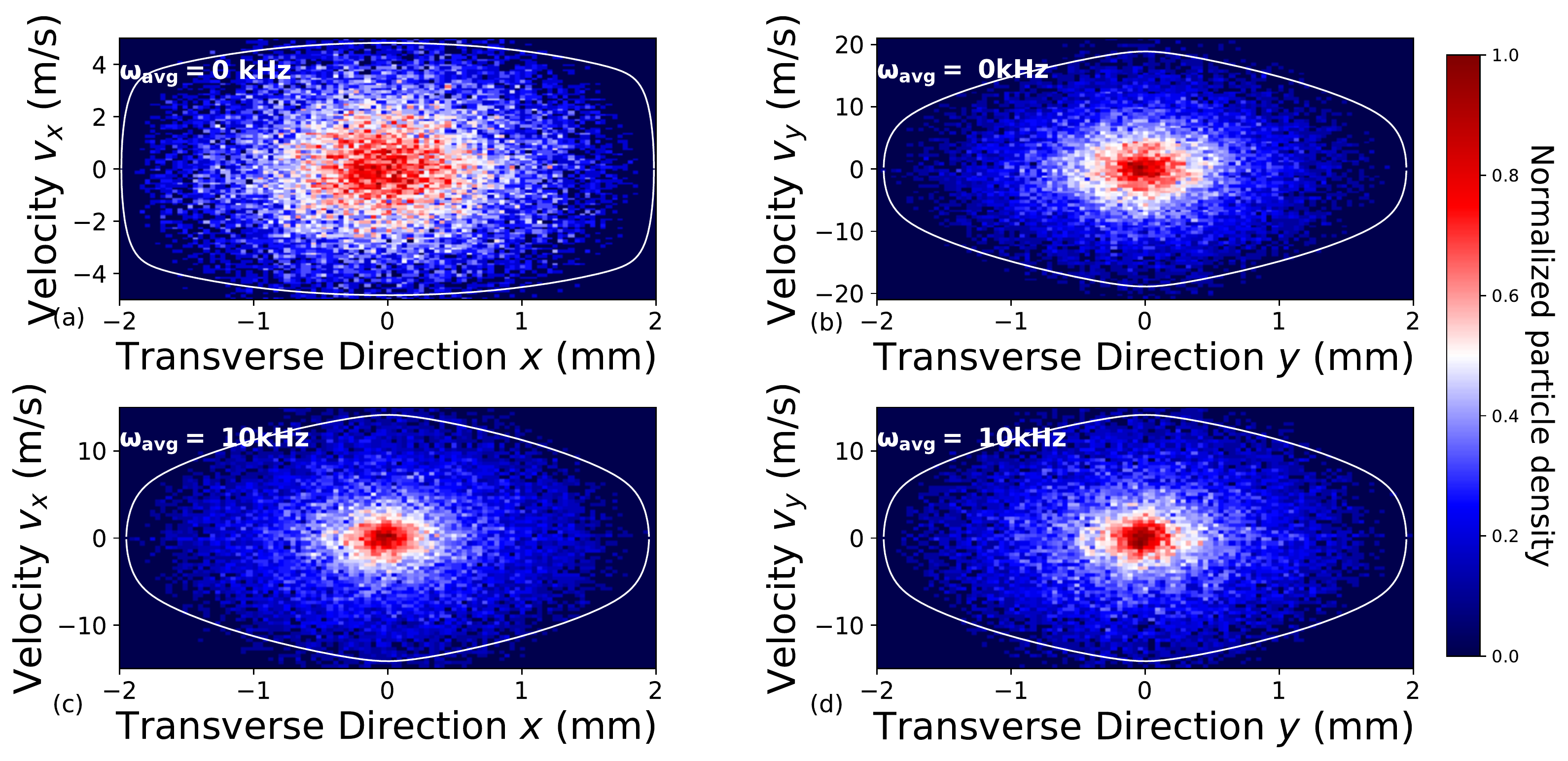}
\centering
\caption{Phase-space distributions of molecules in the transverse directions (a), (c) $x$ and (b),(d) $y$ obtained from 3D numerical trajectory simulations. Panels (a) and (b) correspond to $\omega_\mathrm{avg}=0$, while (c) and (d) illustrate the case $\omega_\mathrm{avg} = 10$ kHz. White lines represent the calculated separatrices of the phase-stable regions for each case.}
\label{fig:psa1}
\end{figure}    

\indent To illustrate the phase-space stability of the new deceleration method, we show in~\Fref{fig:psa} (a)-(e) 2D longitudinal phase-space distributions of particles in a single trap at the end of the decelerator for decelerations ranging from 0.25~km/s$^2$ to 55.8~m/s$^2$ corresponding to final velocities in the range 449~m/s to 50~m/s. Panel (f) corresponds to a deceleration of -27.6~m/s$^2$, i.e., an acceleration to a final velocity of 550~m/s. A normalized particle density is shown as a color map, the calculated separatrices delineating the phase-stable regions are represented by white solid lines. As evidenced by~\Fref{fig:psa}, the phase-space volume transported through the decelerator decreases with increasing deceleration, as is also the case with previous implementations of Stark- and Zeeman decelerators~\cite{meerakker05c,wiederkehr10a}.  

Additionally, we explored the 2D phase-space dynamics in both transverse directions for the case of a stationary transverse trap ($\omega_\mathrm{avg}$ = 0 kHz) and a trap rotating at $\omega_\mathrm{avg}$ = 10 kHz. Comparatively long simulation times of 50 ms were chosen to allow the particles to fully explore the phase space. \Fref{fig:psa1} illustrates transverse phase-space distributions extracted from the trajectory simulations. The density of the particles is represented by a color map. \Fref{fig:psa1} (a) and (b) shows the phase-space distributions of the particles in the $x$ and $y$ for $\omega_\mathrm{avg}$ = 0 kHz while panels (c) and (d) correspond to the case $\omega_\mathrm{avg}$ = 10 kHz. Separatrices are indicated by white lines. We infer a factor of $\sim$2 difference in transverse phase-space acceptance between the two cases. These results are in agreement with transverse magnetic field distribution outlined in~\Sref{sec:transstab},~\Fref{fig:modbfield} and~\Fref{fig:anharm}, where it was shown that stable trapping regions exist for both $\omega_\mathrm{avg}$=0 kHz and 10 kHz along the $y$ direction. Along the $x$ direction at a rotational frequency $\omega_\mathrm{avg}$=0 kHz, the phase-space distribution shows free-flight characteristics due to very small confinement. At a rotational frequency  $\omega_\mathrm{avg}$=10 kHz, there is no appreciable difference in phase-space structure between the $x$ and $y$ directions as the particles are trapped in the same time-averaged potential along both directions during the deceleration process. 
\begin{figure}[t]
\includegraphics[width=0.99\textwidth]{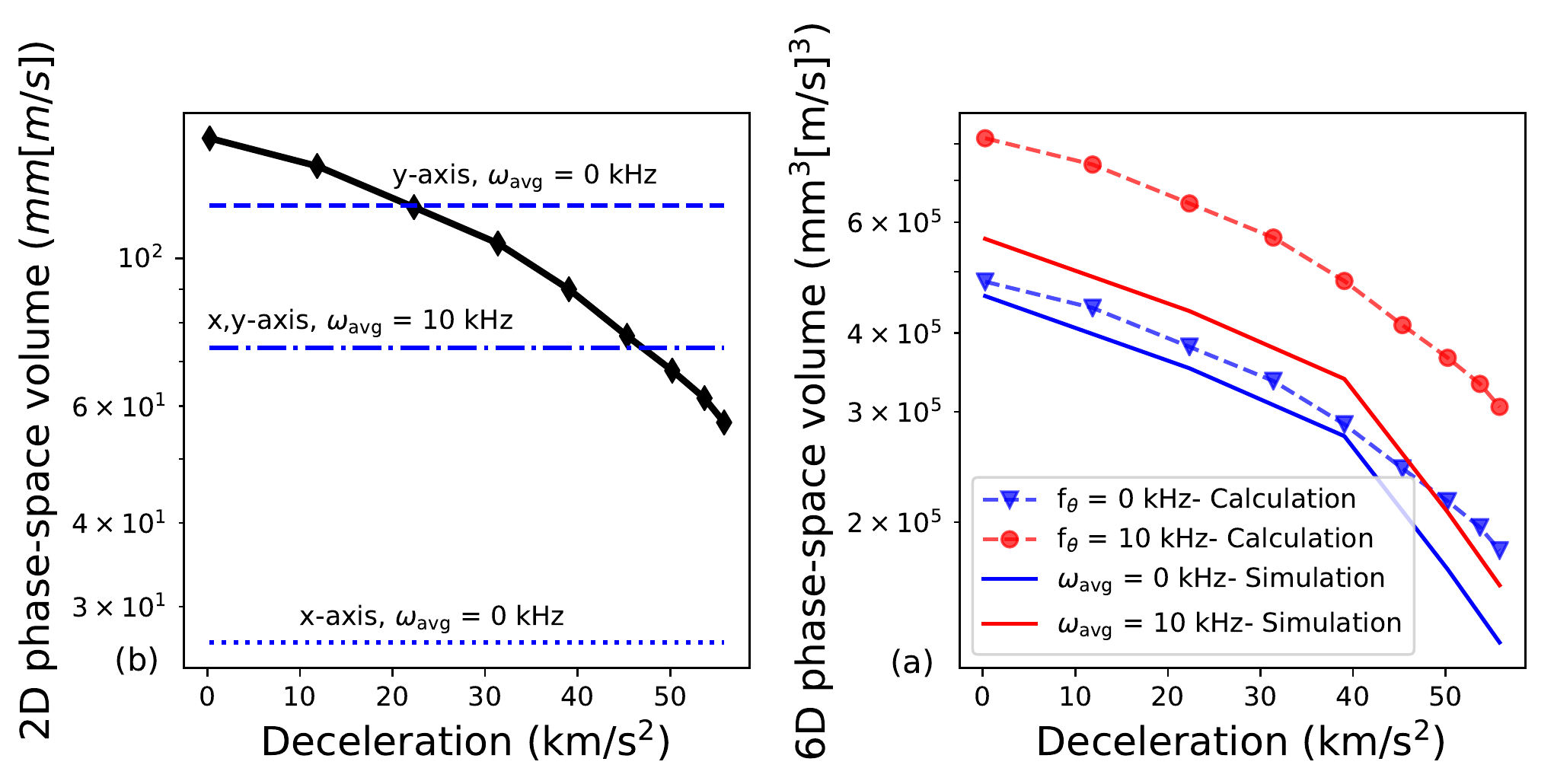}
\centering
\caption{Phase-space acceptance of the traveling-wave Zeeman decelerator as a function of the deceleration. (a) Calculated 2D phase-space acceptance for the longitudinal (black) and transverse (blue) directions with $\omega_\mathrm{avg}$ = 0 and 10 kHz. (b)  6D phase-space acceptance for $\omega_\mathrm{avg}$ = 10 kHz (dashed blue line) and 0 kHz (dashed red line) estimated from a product of 2D acceptance volumes along each coordinate. Results from the corresponding full 6D numerical trajectory simulations are shown by full lines. See text for details.}
\label{fig:psa2}
\end{figure}  
\\
In addition, the complete 6D phase-space acceptance was explored numerically for a range of decelerations. Results of these calculations are shown in~\Fref{fig:psa2} (a). The 2D longitudinal phase-space acceptance is shown by the black line, while the transverse phase-space acceptances both for a stationary ($\omega_\mathrm{avg}$ = 0 kHz) and rotating ($\omega_\mathrm{avg}$ = 10 kHz) trap are indicated by the blue lines. The full 6D phase-space acceptance for a range of decelerations is shown in~\Fref{fig:psa2} (b). An upper limit of the complete 6D phase-space volume is approximated by a product of the 2D volumes along each coordinate. The dashed traces in~\Fref{fig:psa2} (b) show the thus calculated maximum 6D phase-space acceptance for  $\omega_\mathrm{avg}$ = 10 kHz (blue dashed line) and $\omega_\mathrm{avg}$ = 0 kHz (red dashed line). Additionally, the full 6D phase-space volume at the end of the deceleration region was extracted from the numerical trajectory simulations (blue and red full lines in ~\Fref{fig:psa2} (b)). The difference between the phase-space acceptances estimated from the product of the 2D acceptances and the ones obtained from the trajectory simulations is attributed to transverse instabilities of trajectories far from the centre of the trap.

\noindent The maximum phase-space acceptance volumes calculated for the highest deceleration (55.8 km\/s$^{2}$ corresponding to the final velocity of around 50 m\/s) are $8\cdot10^4$ $\mathrm{mm^3~m/s^3}$ for $\omega_\mathrm{avg}$ = 10 kHz and $5\cdot10^4$  $\mathrm{mm^3~m/s^3}$ for $\omega_\mathrm{avg}$ = 0 kHz. For comparison, the maximum phase-space acceptance volume of a typical Stark decelerator at a final velocity of 50 m/s is $8\cdot10^3$  $\mathrm{mm^3~m/s^3}$~\cite{zhang16a,Shyur18a}, and $2\cdot10^7$  $\mathrm{mm^3~m/s^3}$ for the Zeeman decelerator reported in~\cite{Lavert_Ofir11a}. The phase-space acceptance of the current assembly could be further enhanced by increasing the currents supplied to the coils. 

\section{Conclusions}
We have developed the theory describing the working principle of a novel traveling-trap Zeeman decelerator including analytical expressions for the magnetic fields generated from a double-helix coil geometry in one dimension. Two independent parameters governing the dynamics of the moving magnetic traps were introduced. The transverse stability of the decelerator was explored in detail. It was shown that trap stability in relation to the rotation frequency $\omega_\mathrm{avg}$ enables the selectivele trapping of molecules according to the magnetic-dipole-moment-to-mass ratio and the current amplitude supplied to the decelerator. 
The deceleration efficiency is limited by the deceleration applied to the molecules which is an intrinsic property of decelerators based on conservative forces. 

\section*{Acknowledgments}
Funding from the Swiss National Science Foundation, grant nr. 200020\_175533, and the University of Basel is acknowledged. D. Z. acknowledges the financial support from Freiwillige Akademische Gesellschaft (FAG) Basel, the Research Fund for Junior Researchers of the University of Basel and National Key R\&D Program of China, grant nr. 2019YFA0307701.

\section{Supplementary information}

The magnetic field inside a single right-hand oriented helix can be calculated using the Biot-Savart law:
\begin{equation}
    \mathbf{B}(\mathbf{r}) = \frac{\mu_0}{4\pi}\int_C\frac{Id\mathbf{l}\times\mathbf{r'}}{|r'|^3}.
\end{equation}
Parameterizing a right-handed helix in cylindrical coordinates 
\begin{equation}
    l(\theta) = (R \cos\theta, \ R\sin\theta,\ \frac{\lambda\theta}{2\pi}),
\end{equation}
and substituting them into the Biot-Savart law gives expressions for the components of the magnetic field inside a single helix as
\begin{align}
    \begin{split}
    \label{eqn:ch4_bf1x}
      B_x =\frac{\mu_0I}{4\pi}\int \frac{\lambda  R \sin \theta +R \cos \theta  (2 \pi  z-\theta  \lambda )-\lambda 
   y}{2 \pi  \left((x-R \cos \theta )^2+(y-R \sin \theta )^2+\left(z-\frac{\theta 
   \lambda }{2 \pi }\right)^2\right)^{3/2}}d\theta,
    \end{split}\\
    \begin{split}
     \label{eqn:ch4_bf1y}
        B_y =\frac{\mu_0I}{4\pi}\int \frac{-\lambda  R \cos \theta +R \sin \theta  (2 \pi  z-\theta  \lambda )+\lambda 
   x}{2 \pi  \left((x-R \cos \theta )^2+(y-R \sin \theta )^2+\left(z-\frac{\theta 
   \lambda }{2 \pi }\right)^2\right)^{3/2}}d\theta,
    \end{split}\\
    \begin{split}
     \label{eqn:ch4_bf1z}
        B_z =\frac{\mu_0I}{4\pi}\int  \frac{R (R-x \cos \theta -y \sin \theta )}{\left((x-R \cos \theta )^2+(y-R \sin
   \theta )^2+\left(z-\frac{\theta  \lambda }{2 \pi }\right)^2\right)^{3/2}}d\theta,
    \end{split}
\end{align}
where $\theta$ is the parameterized angle in cylindrical coordinates, $R$ is the radius of the helix, $\lambda$ is the periodicity of the helix and $I$ is the current supplied to the helix. 
From equations (\ref{eqn:ch4_bf1x})-(\ref{eqn:ch4_bf1z}), the magnetic field components along $z$ axis can be calculated. Setting $x = 0 $ and $y = 0$, equations \ref{eqn:ch4_bf1x}-\ref{eqn:ch4_bf1z} reduce to:
\begin{equation}
    B_x = \frac{\mu_0 I}{4 \pi}\int \frac{\lambda  R \sin \theta +R \cos \theta  (2 \pi 
   z-\theta  \lambda )}{2 \pi  \left(R^2+\left(z-\frac{\theta  \lambda }{2 \pi
   }\right)^2\right)^{3/2}}d\theta,
    \label{eqn:1dbf1}
\end{equation}
\begin{equation}
    B_y = \frac{\mu_0 I}{4 \pi}\int \frac{-\lambda  R \cos \theta +R \sin \theta  (2 \pi 
   z-\theta  \lambda )}{2 \pi  \left(R^2+\left(z-\frac{\theta  \lambda }{2 \pi
   }\right)^2\right)^{3/2}}d\theta,
   \label{eqn:1dbf2}
\end{equation}
\begin{equation}
    B_z = \frac{\mu_0 I}{4 \pi}\int \frac{R^2}{\left(R^2+\left(z-\frac{\theta  \lambda }{2\pi
   }\right)^2\right)^{3/2}}d\theta.
   \label{eqn:1dbf3}
\end{equation}
With the use of the integral identities:
\begin{equation}
    \int_{-\infty}^{\infty}\frac{\sin\big(\alpha(\xi-\eta)\big)}{(r^2+\eta^2)^{3/2}}d\eta = \frac{2\alpha}{r} K_1(\alpha r)\sin(\alpha\xi),
\end{equation}
\begin{equation}
    \int_{-\infty}^{\infty}\frac{\cos\big(\alpha(\xi-\eta)\big)}{(r^2+\eta^2)^{3/2}}d\eta = \frac{2\alpha}{r} K_1(\alpha r)\cos(\alpha\xi),
\end{equation}
\begin{equation}
    \int_{-\infty}^{\infty}\frac{\eta\sin\big(\alpha(\xi-\eta)\big)}{(r^2+\eta^2)^{3/2}}d\eta = -2\alpha K_0(\alpha r)\cos(\alpha\xi),
\end{equation}
\begin{equation}
    \int_{-\infty}^{\infty}\frac{\eta\cos\big(\alpha(\xi-\eta)\big)}{(r^2+\eta^2)^{3/2}}d\eta = 2\alpha K_0(\alpha r)\sin(\alpha\xi),
\end{equation}
\begin{equation}
    \int_{-\infty}^{\infty}\frac{1}{(r^2+\eta^2)^{3/2}}d\eta = \frac{2}{r^2}
\end{equation}
and setting $r = R$, $\alpha = \frac{2\pi}{\lambda}$, $\xi  =z$, the on-axis magnetic field components evaluate to:
\begin{equation}
    B_x = \frac{\mu_0I}{\lambda}I\Big[K_1(\frac{2\pi R}{\lambda})+\frac{2 \pi R}{\lambda}K_0(\frac{2\pi R}{\lambda})\Big]\sin(\frac{2\pi}{\lambda}z),
    \label{eqn:tf1}
\end{equation}
\begin{equation}
    B_y = \frac{\mu_0I}{\lambda}I\Big[K_1(\frac{2\pi R}{\lambda})+\frac{2 \pi R}{\lambda}K_0(\frac{2\pi R}{\lambda})\Big]\cos(\frac{2\pi}{\lambda}z),
    \label{eqn:tf2}
\end{equation}
\begin{equation}
    B_z = \frac{\mu_0 I}{\lambda}
    \label{eqn:tf3},
\end{equation}
where $K_n(x)$ are the modified Bessel functions of the second kind.
Similarly, the magnetic field inside a single left-hand-oriented helix is:

\begin{equation}
     B_x = -\frac{\mu_0 I}{4 \pi}\int \frac{\lambda R \sin \theta +R \cos \theta (2
   \pi  z-\theta  \lambda )+\lambda y}{2 \pi  \left((x-R \cos
   \theta )^2+(R \sin \theta
   +y)^2+\left(z-\frac{\theta  \lambda }{2 \pi
   }\right)^2\right)^{3/2}}d\theta
\end{equation}
\begin{equation}
     B_y =\frac{\mu_0I}{4\pi}\int\frac{-\lambda  R \cos \theta+R \sin \theta (2 \pi 
   z-\theta  \lambda )+\lambda  x}{2 \pi  \left((x-R \cos
   \theta )^2+(R \sin \theta
   +y)^2+\left(z-\frac{\theta  \lambda }{2 \pi
   }\right)^2\right)^{3/2}}d\theta,
\end{equation}

\begin{equation}
     B_z =-\frac{\mu_0I}{4\pi}\int\frac{R (R-x \cos \theta+y \sin \theta
  )}{\left((x-R \cos \theta )^2+(R \sin \theta
   +y)^2+\left(z-\frac{\theta  \lambda }{2 \pi
   }\right)^2\right)^{3/2}}d\theta.
\end{equation}
Taking $x=0$ and $y=0$ reduces to:
\begin{equation}
    B_x = -\frac{\mu_0 I}{4 \pi}\int \frac{\lambda  R \sin \theta +R \cos \theta  (2 \pi 
   z-\theta  \lambda )}{2 \pi  \left(R^2+\left(z-\frac{\theta  \lambda }{2 \pi
   }\right)^2\right)^{3/2}}d\theta,
    \label{eqn:1dbf4}
\end{equation}
\begin{equation}
    B_y = \frac{\mu_0 I}{4 \pi}\int \frac{-\lambda  R \cos \theta +R \sin \theta  (2 \pi 
   z-\theta  \lambda )}{2 \pi  \left(R^2+\left(z-\frac{\theta  \lambda }{2 \pi
   }\right)^2\right)^{3/2}}d\theta,
   \label{eqn:1dbf5}
\end{equation}
\begin{equation}
    B_z = -\frac{\mu_0 I}{4 \pi}\int \frac{R^2}{\left(R^2+\left(z-\frac{\theta  \lambda }{2\pi
   }\right)^2\right)^{3/2}}d\theta.
   \label{eqn:1dbf6}
\end{equation}
In similar fashion as for the right-hand helix, one obtains the expressions for the on-axis magnetic field components for the left-hand oriented helix:
\begin{equation}
    B_x = -\frac{\mu_0I}{\lambda}\Big[K_1(\frac{2\pi R}{\lambda})+\frac{2 \pi R}{\lambda}K_0(\frac{2\pi R}{\lambda})\Big]\sin(\frac{2\pi}{\lambda}z),
    \label{eqn:tf4}
\end{equation}
\begin{equation}
    B_y = \frac{\mu_0I}{\lambda}\Big[K_1(\frac{2\pi R}{\lambda})+\frac{2 \pi R}{\lambda}K_0(\frac{2\pi R}{\lambda})\Big]\cos(\frac{2\pi}{\lambda}z),
    \label{eqn:tf5}
\end{equation}
\begin{equation}
    B_z = -\frac{\mu_0 I}{\lambda}.
    \label{eqn:tf6}
\end{equation}
In these calculations, the approximation of an infinitely extended helix was used.

From~\Eref{eqn:tf1}-\Eref{eqn:tf3}) and~\Eref{eqn:tf4}-\Eref{eqn:tf6}, the general form of the on-axis magnetic field for both right- and left-hand-oriented helix is given and the parameters $a_+,a_-,a_{z+}$ and $a_{z-}$ given in the main text can be obtained:
\begin{equation}
    a_+ =\frac{\mu_0I}{\lambda}\Big[K_1(\frac{2\pi R_1}{\lambda})+\frac{2 \pi R_1}{\lambda}K_0(\frac{2\pi R_1}{\lambda})\Big],
\end{equation}
\begin{equation}
    a_- = \frac{\mu_0I}{\lambda}\Big[K_1(\frac{2\pi R_2}{\lambda})+\frac{2 \pi R_2}{\lambda}K_0(\frac{2\pi R_2}{\lambda})\Big],
\end{equation}
\begin{equation}
    a_{+z} = \frac{\mu_0}{\lambda},
\end{equation}
\begin{equation}
    a_{-z} = -\frac{\mu_0}{\lambda},
\end{equation}
where $R_1$ and $R_2$ are the radii of the right- and left-hand-oriented helix, respectively. 

\section*{References}
\bibliographystyle{iopart-num}
\bibliography{bib,ref}

\providecommand{\newblock}{}
\begin{thebibliography}{10}
\expandafter\ifx\csname url\endcsname\relax
  \def\url#1{{\tt #1}}\fi
\expandafter\ifx\csname urlprefix\endcsname\relax\def\urlprefix{URL }\fi
\providecommand{\eprint}[2][]{\url{#2}}

\bibitem{carr09a}
Carr L~D, DeMille D, Krems R~V and Ye J 2009 {\em New J. Phys.\/} {\bf 11}
  055049

\bibitem{balakrishnan16a}
Balakrishnan N 2016 {\em J Chem. Phys.\/} {\bf 145} 150901

\bibitem{deMille17a}
DeMille D, Doyle J~M and Sushkov A~O 2017 {\em Science\/} {\bf 357} 990

\bibitem{tarbutt18a}
Tarbutt M~R 2018 {\em Contemp. Phys.\/} {\bf 59} 356

\bibitem{meerakker12a}
\mbox{van de} Meerakker S~Y~T, Bethlem H~L, Vanhaecke N and Meijer G 2012 {\em
  Chem. Rev.\/} {\bf 112} 4828

\bibitem{narevicius12a}
Narevicius E and Raizen M~G 2012 {\em Chem. Rev.\/} {\bf 112} 4879

\bibitem{jin12a}
Jin D~S and Ye J 2012 {\em Chem. Rev.\/} {\bf 112} 4801

\bibitem{Bohn17a}
Bohn J~L, Rey A~M and Ye J 2017 {\em Science\/} {\bf 357} 1002--1010

\bibitem{safronova18a}
Safronova M~S, Budker D, DeMille D, Kimball D~F~J, Derevianko A and Clark C~W
  2018 {\em Rev. Mod. Phys.\/} {\bf 90}(2) 025008

\bibitem{chupp19a}
Chupp T~E, Fierlinger P, Ramsey-Musolf M~J and Singh J~T 2019 {\em Rev. Mod.
  Phys.\/} {\bf 91}(1) 015001

\bibitem{krems09a}
Krems R~V, Stwalley W~C and Friedrich B (eds) 2009 {\em Cold Molecules: Theory,
  Experiment, Applications\/} (Boca Raton: CRC Press)

\bibitem{georgescu14a}
Georgescu I~M, Ashhab S and Nori F 2014 {\em Rev. Mod. Phys.\/} {\bf 86}(1)
  153--185

\bibitem{mcardle20a}
McArdle S, Endo S, Aspuru-Guzik A, Benjamin S~C and Yuan X 2020 {\em Rev. Mod.
  Phys.\/} {\bf 92}(1) 015003

\bibitem{Kirste12a}
Kirste M, Wang X, Schewe H~C, Meijer G, Liu K, \mbox{van der Avoird} A, Janssen
  L~M~C, Gubbels K~B, Groenenboom G~C and van~de Meerakker S~Y~T 2012 {\em
  Science\/} {\bf 338} 1060

\bibitem{vogel14a}
Vogels S~N, Onvlee J, von Zastrow A, Groenenboom G~C, van~der Avoird A and
  van~de Meerakker S~Y~T 2014 {\em Phys. Rev. Lett.\/} {\bf 113}(26) 263202

\bibitem{Vogels15a}
Vogels S~N, Onvlee J, Chefdeville S, van~der Avoird A, Groenenboom G~C and
  van~de Meerakker S~Y~T 2015 {\em Science\/} {\bf 350} 787

\bibitem{Akerman15a}
Akerman N, Karpov M, David L, Lavert-Ofir E, Narevicius J and Narevicius E 2015
  {\em New J. Phys\/} {\bf 17} 065015

\bibitem{gao18a}
Gao Z, Karman T, Vogels S~N, Besemer M, van~der Avoird A, Groenenboom G~C and
  van~de Meerakker S~Y~T 2018 {\em Nat. Chem.\/} {\bf 10} 469

\bibitem{vogels18a}
Vogels S~N, Karman T, Kłos J, Besemer M, Onvlee J, van~der Avoird A,
  Groenenboom G~C and van~de Meerakker S~Y~T 2018 {\em Nat. Chem.\/} {\bf 10}
  435

\bibitem{segev19a}
Segev Y, Pitzer M, Karpov M, Akerman N, Narevicius J and Narevicius E 2019 {\em
  Nature\/} {\bf 572} 189

\bibitem{vanhaecke07a}
Vanhaecke N, Meier U, Andrist M, Meier B~H and Merkt F 2007 {\em Phys.
  Rev.~A\/} {\bf 75} 031402

\bibitem{Narevicius07a}
Narevicius E, Parthey C~G, Libson A, Narevicius J, Chavez I, Even U and Raizen
  M~G 2007 {\em New J Phys.\/} {\bf 9} 358

\bibitem{hogan07a}
Hogan S~D, Sprecher D, Andrist M, Vanhaecke N and Merkt F 2007 {\em Phys.
  Rev.~A\/} {\bf 76} 023412

\bibitem{hogan08a}
Hogan S~D, Wiederkehr A~W, Schmutz H and Merkt F 2008 {\em Phys. Rev. Lett.\/}
  {\bf 101} 143001

\bibitem{wiederkehr11a}
Wiederkehr A~W, Motsch M, Hogan S~D, Andrist M, Schmutz H, Lambillotte B, Agner
  J~A and Merkt F 2011 {\em J Chem. Phys.\/} {\bf 135} 214202

\bibitem{wiederkehr12a}
Wiederkehr A, Schmutz H, Motsch M and Merkt F 2012 {\em Mol. Phys.\/} {\bf 110}
  1807

\bibitem{trimeche11a}
Trimeche A, Bera M~N, Cromi\'{e}res J~P, Robert J and Vanhaecke N 2011 {\em
  Eur. Phys. J. D\/} {\bf 65} 263

\bibitem{Lavert_Ofir11a}
Lavert-Ofir E, Gersten S, Henson A~B, Shani I, David L, Narevicius J and
  Narevicius E 2011 {\em New J Phys.\/} {\bf 13} 103030

\bibitem{motsch14a}
Motsch M, Jansen P, Agner J~A, Schmutz H and Merkt F 2014 {\em Phys. Rev. A\/}
  {\bf 89}(4) 043420

\bibitem{liu15a}
Liu Y, Zhou S, Zhong W, Djuricanin P and Momose T 2015 {\em Phys. Rev. A\/}
  {\bf 91} 021403(R)

\bibitem{cremers17a}
Cremers T, Chefdeville S, Janssen N, Sweers E, Koot S, Claus P and van~de
  Meerakker S~Y~T 2017 {\em Phys. Rev. A\/} {\bf 95} 043415

\bibitem{akerman17a}
Akerman N, Karpov M, Segev Y, Bibelnik N, Narevicius J and Narevicius E 2017
  {\em Phys. Rev. Lett.\/} {\bf 119}(7) 073204

\bibitem{semeria18a}
Semeria L, Jansen P, Clausen G, Agner J~A, Schmutz H and Merkt F 2018 {\em
  Phys. Rev. A\/} {\bf 98} 062518

\bibitem{cremers18a}
Cremers T, Chefdeville S, Plomp V, Janssen N, Sweers E and van~de Meerakker
  S~Y~T 2018 {\em Phys. Rev. A\/} {\bf 98}(3) 033406

\bibitem{mcard18a}
{McArd} L~A, Mizouri A, Walker P~A, Singh V, Krohn U, Hinds E~A and Carty D
  2018 {\em {arXiv}:1807.10648 [physics]\/} (\textit{Preprint}
  \eprint{1807.10648})

\bibitem{plomp19a}
Plomp V, Gao Z, Cremers T and van~de Meerakker S~Y~T 2019 {\em Phys. Rev. A\/}
  {\bf 99}(3) 033417

\bibitem{cremers19a}
Cremers T, Janssen N, Sweers E and van~de Meerakker S~Y~T 2019 {\em Rev. Sci.
  Instrum.\/} {\bf 90} 013104

\bibitem{hogan11a}
Hogan S~D, Motsch M and Merkt F 2011 {\em Phys. Chem. Chem. Phys.\/} {\bf 13}
  18705

\bibitem{meek08a}
Meek S~A, Bethlem H~L, Conrad H and Meijer G 2008 {\em Phys. Rev. Lett.\/} {\bf
  100}(15) 153003

\bibitem{meek09a}
Meek S~A, Conrad H and Meijer G 2009 {\em Science\/} {\bf 324} 1699

\bibitem{meek09b}
Meek S~A, Conrad H and Meijer G 2009 {\em New. J. Phys\/} {\bf 11} 055024

\bibitem{osterwalder10a}
Osterwalder A, Meek S~A, Hammer G, Haak H and Meijer G 2010 {\em Phys. Rev.
  A\/} {\bf 81} 051401

\bibitem{meek11a}
Meek S~A, Parsons M~F, Heyne G, Platschkowski V, Haak H, Meijer G and
  Osterwalder A 2011 {\em Rev. Sci. Instru.\/} {\bf 82} 093108

\bibitem{bulleid12a}
Bulleid N~E, Hendricks R~J, Hinds E~A, Meek S~A, Meijer G, Osterwalder A and
  Tarbutt M~R 2012 {\em Phys. Rev. A\/} {\bf 86}(2) 021404

\bibitem{damjanovic21b}
Damjanovi\'c T, Willitsch S, Vanhaecke N, Haak H, Meijer G, Cromi\'eres J and
  Zhang D 2021 {\em Manuscript submitted for publication.\/}

\bibitem{meerakker05c}
van~de Meerakker S~Y~T, Vanhaecke N, Bethlem H~L and Meijer G 2005 {\em Phys.
  Rev. A\/} {\bf 71}(5) 053409

\bibitem{wiederkehr10a}
Wiederkehr A~W, Hogan S~D and Merkt F 2010 {\em Phys. Rev. A\/} {\bf 82}(4)
  043428

\bibitem{brown03a}
Brown J~M and Carrington A 2003 {\em Rotational Spectroscopy of Diatomic
  Molecules\/} (Cambridge University Press)

\bibitem{haas19a}
Haas D {\em Towards hybrid trapping of cold molecules and cold molecular
  ions\/} Ph.D. thesis University of Basel

\bibitem{lam2015numba15a}
Lam S~K, Pitrou A and Seibert S 2015 Numba: A llvm-based python jit compiler
  {\em Proceedings of the Second Workshop on the LLVM Compiler Infrastructure
  in HPC\/} pp 1--6

\bibitem{zhang16a}
Zhang D, Meijer G and Vanhaecke N 2016 {\em Phys. Rev. A\/} {\bf 93}(2) 023408

\bibitem{Shyur18a}
Shyur Y, Bossert J~A and Lewandowski H~J 2018 {\em J. Phys.s B: At. Mol.
  Opt.\/} {\bf 51} 165101

\end{thebibliography}

\end{document}